\documentclass[conference]{IEEEtran}
\IEEEoverridecommandlockouts
\usepackage{cite}
\usepackage{amsmath,amssymb,amsfonts}
\usepackage{algorithmic}
\usepackage{graphicx}
\usepackage{textcomp}
\usepackage{xcolor}
\usepackage{tabularx}
\usepackage{todonotes}
\newtheorem{theorem}{Theorem}

\newcommand{\interactions}{\mathbf{R}}
\newcommand{\predictions}{\mathbf{\widehat{R}}}
\newcommand{\reclist}{\mathcal{R}^{(i)}}
\newcommand{\mean}{\bar{\mu}}
\newcommand{\covariance}{\mathbf{\Sigma}}
\newcommand{\sd}{\bar{\sigma}}
\newcommand{\liked}{\mathcal{I}_{i}}
\newcommand{\notliked}{\mathcal{J}_{i}}
\newcommand{\other}{\mathcal{O}_{j}}
\newcommand{\reference}{\mathcal{R}}
\newcommand{\remaining}{\mathcal{S}}
\newcommand{\correlation}{\mathbf{C}}
\newcommand{\similarity}{\mathbf{S}}
\newcommand{\neighbourhood}{\mathcal{D}^{(k)}}
\newcommand{\users}{\mathbf{P}}
\newcommand{\items}{\mathbf{G}}
\newcommand{\userbias}{\bar{p}}
\newcommand{\itembias}{\bar{g}}
\newcommand{\centered}{\mathbf{X}}
\newcommand{\coefmat}{\mathbf{B}}
\newcommand{\coefvec}{\bar{\beta}}
\newcommand{\leftsing}{\mathbf{U}}
\newcommand{\valsing}{\mathbf{\Psi}}
\newcommand{\rightsing}{\mathbf{V}}
\newcommand{\sign}{\psi}
\newcommand{\projection}{\mathbf{T}}
\newcommand{\usersmissing}{\widetilde{\mathbf{P}}}

\def\BibTe\centered{{\rm B\kern-.05em{\sc i\kern-.025em b}\kern-.08em
    T\kern-.1667em\lower.7ex\hbox{E}\kern-.125em\centered}}
\begin{document}

\title{New Recommendation Algorithm for Implicit Data Motivated by the Multivariate Normal Distribution}

\author{\IEEEauthorblockN{Markus Viljanen, Tapio Pahikkala}
\IEEEauthorblockA{\textit{Department of Future Technologies} \\
\textit{University of Turku}\\
Turku, Finland
}
}

\maketitle

\begin{abstract}

The goal of recommender systems is to help users find useful items from a large catalog of items by producing a list of item recommendations for every user. Data sets based on implicit data collection have a number of special characteristics. The user and item interaction matrix is often complete, i.e. every user and item pair has an interaction value or zero for no interaction, and the goal is to rank the items for every user. This study presents a simple new algorithm for implicit data that matches or outperforms baselines in accuracy. The algorithm can be motivated intuitively
by the Multivariate Normal Distribution (MVN), where have a closed form expression for the ranking of non-interactions given user's interactions. The main difference to kNN and SVD baselines is that predictions are carried out using only the known interactions. Modified baselines with this trick have a better accuracy, however it also results in simpler models with fewer hyperparameters for implicit data. Our results suggest that this idea should used in Top-N recommendation with small seed sizes and the MVN is a simple way to do so.

\end{abstract}


\section{Introduction}

Recommender systems are algorithms that provide suggestions of new items to users. These suggestions enable users to find useful items from a large catalog of items. Research into recommender systems is a large academic field that has found many successful applications \cite{review1}\cite{review2}: movies, music, books, products, etc. Recommender systems have also become popular in the industry, where the goal often is to increase sales and user engagement \cite{ecommerence}. A recommender system is fitted to a training set of past interactions and evaluated on a test set of new interactions. Each user has some past interactions as a seed on which the recommendations are based. A large data set of historical user and item interactions enables the development of most effective algorithms. 

The data sets can be classified as either explicit or implicit \cite{handbook}, based on whether interactions are based on explicit mentions of items by the users or implicit data collection of users interacting with items. Explicit data sets typically contains missing values, for example users have provided some movie ratings and the goal is to predict the missing ratings. Implicit data sets in contrast are often complete, for example each interaction is one or zero denoting whether a given user has watched a given movie and the goal is to rank the not yet watched movies for every user. In this study, we focus data sets where every entry in the interaction matrix has a value that contains an interaction strength or a zero for no interaction. This setting is most naturally implied by an implicit data set. The task is to produce a list of item recommendations to a user, which is is called top-N recommendation \cite{topn0}\cite{topn1}\cite{topn2}\cite{topn3}. This is used in many commercial recommender systems and is often the ultimate use case, regardless of whether the recommender system predicts missing values or rankings \cite{mostrelevant}\cite{evaluation1}. 

Most popular algorithms can be classified into two categories: collaborative filtering \cite{collaborative} and content based \cite{content}. In collaborative filtering, recommendations are based on how users and items have interacted, without any additional knowledge of the users or the items. In content based filtering, information about the users and items is used to recommend suitable items to the user. These algorithms have different strengths and they can be merged to produce so called hybrid recommenders \cite{hybrid}. Model based collaborative filtering typically produces more accurate recommendations, unless predictions are required for new items with very few or no interactions \cite{better1}\cite{better2}. This is also referred to as the cold-start problem \cite{coldstart}. While academic research has typically focused on new methods that improve the accuracy in some task, there is increasing awareness that comparing models by their accuracy may not always be optimal for perceived utility \cite{evaluation2}\cite{evaluation3}\cite{evaluation4}. In this study, we focus on collaborative filtering and view the content based approaches as complementary. We compare methods primarily by their accuracy, but also briefly discuss qualitative results and the effect of item popularity.

Our main contribution is a new recommendation algorithm for implicit data sets that outperforms baselines in accuracy and is much simpler. The algorithm can be motivated intuitively by the Multivariate Normal Distribution (MVN) where user's rating vector is an outcome. The idea is to think of the non-interaction values for each user as 'missing data' to be predicted. We train the model simply by computing the mean vector and the covariance matrix from the user and item interaction matrix, then predict the Top-N ranking for each user as the conditional mean of non-interaction values given the interaction values. The predictions are calculated directly from data by a closed form expression. We show that our model differs from standard baselines in one important way: the non-interactions are not taken into account when the interactions of every user are predicted. It is possible to modify the standard baselines accordingly and the resulting models have a much better accuracy. As an additional benefit, it is not necessary to limit the number of latent dimensions or perform exhaustive searches for the regularization parameter.

\section{Related Work}

Collaborative filtering algorithms can be generally classified into neighborhood and model-based methods.\cite{explicit_advances}. The neighbourhood methods are one of the earliest and remain the most popular. User-based predict unknown ratings based on the ratings of similar users, whereas item-based predict unknown ratings based on known ratings made by the same user on similar items. Many early approaches were user-based \cite{nn_user}, but the analogous item-based approaches \cite{nn_item1}\cite{nn_item2} have become more popular as they typically offer better scalability and higher accuracy \cite{nn_better}. The model-based methods are very diverse, they for example include Latent Dirichlet Allocation \cite{latent1}, Latent Semantic Analysis \cite{latent2}, Restricted Boltzmann machines \cite{latent3}, and Neural Networks \cite{latent4}.  Matrix factorization models that resemble the Singular Value Decomposition (SVD) have gained the most popularity, thanks to their high accuracy and scalability \cite{whysvd1}\cite{whysvd2}. Early approaches imputed the missing ratings and used the ordinary SVD \cite{earlysvd}, but modern methods are fitted to the observed ratings only and often add regularization to avoid overfitting \cite{svd1}\cite{svd2}\cite{svd3}\cite{svd4}\cite{svd5}. In the recommender system literature the ordinary SVD is sometimes called PureSVD \cite{topn0}\cite{topn3} and the methods with missing values and regularization is called the SVD \cite{topn2}.

As mentioned before, the data sets can be classified either as either explicit or implicit based on the data collection method \cite{handbook}. Implicit data sets indirectly reflect user opinion through observing their behavior. They are more abundant because users rarely volunteer to generate extensive explicit rating data to use for recommendations \cite{implicit_abundant}. This can make the system more useful in practise \cite{implicit_benefit1}\cite{implicit_benefit2}. In any case, the explicit ratings of users and the items which they implicitly choose to rate tend to correlate \cite{implicit_correlation}. Implicit data sets have a number of special characteristics \cite{topn2}: 1) the data set is often complete because every user has been observed to interact or not interact with every item, 2) there is no explicit negative feedback, for example interactions where the rating signifies a dislike, 3). the feedback is inherently noisy because user's behavior may not perfectly align with their explicitly stated likes and dislikes 4) evaluation has to be based on ranking accuracy metrics such as precision or recall, not on error metrics like the RMSE.

Standard collaborative filtering algorithms can be applied directly to implicit data sets, but they are not necessarily optimal for top-N recommendation \cite{topn0}. Several modifications to the standard algorithms have been presented before in this context, and slightly modified baselines tend to be surprisingly competitive \cite{topn0}\cite{topn1}\cite{topn2}\cite{topn3}. We use the modified baselines in \cite{topn0}. Instead of directly predicting values in the user and item interaction matrix, it also possible to develop methods that predict the relative preferences of users. This can be especially useful in Top-N type prediction tasks and can be seen as an instance of learning-to-rank problem \cite{learningtorank}. There is a similarly long history of such models in collaborative filtering \cite{preference1}\cite{preference2}. Many simple and effective methods can be based on matrix factorization with a special loss function: we mention Bayesian Personalized Ranking (BPR) \cite{bpr} and Logistic matrix factorization \cite{log}.
These methods have efficient implementations for top-N recommendation in the Python packages Implicit \footnote{https://github.com/benfred/implicit} and LightFM \cite{lightfm}, for example. Factorization Machines (FM) are hybrid methods that may include user or item specific features. However, they correspond to matrix factorization when the user and the item feature matrices are identity matrices \cite{fm}, so we can use them as baselines in collaborative filtering tasks.

Doing a fair comparison of the recommendation algorithms is not as easy as one might think. Evaluation is difficult because some algorithms may perform better on different datasets, the goal of the evaluation may differ, and the chosen metrics may also differ \cite{evaluation1}. In addition, finding the optimal hyperparameters and optimization framework to obtain the best performing baselines may require a great deal of effort, and for this reason authors of new methods may have not used optimal baselines \cite{compare1}. There can also be significant differences between identical methods when they are implemented in different software libraries \cite{compare2}. When the studies use a good software implementation with all the methods tuned to the optimum, they tend to produce equally accurate results with standard baselines \cite{compare1} \cite{compare2}.

There is increasing awareness that accuracy alone is not a guarantee of user satisfaction, and the subjective usefulness of a recommender is an important evaluation goal \cite{evaluation2}\cite{evaluation3}\cite{evaluation4}. Accuracy metrics may bias results towards recommending the most popular items instead of personalized recommendations in top-N recommendation. In fact, methods may struggle to deliver results that are personalized in their quest for accuracy \cite{popularity}. Naive algorithms like popularity ranking can sometimes even match the performance of sophisticated algorithms \cite{topn0}. Practical recommender systems could therefore benefit from the ability to adjust the effect of item popularity on the results. The research on this topic is more limited because it is expensive and time-consuming to perform a user study. Online user studies have been performed for example in movies \cite{qualitative1}, news \cite{qualitative2} and games \cite{qualitative3}, where algorithms with different quantitative accuracy had similar user satisfaction. Aside from accuracy and user satisfaction, developers of commercial recommenders should measure the business impact, but academic literature has limited information on the topic \cite{business1}\cite{business2}

\section{Data and Recommendation Problem}

\begin{figure*}
\centerline{\includegraphics[width=\textwidth]{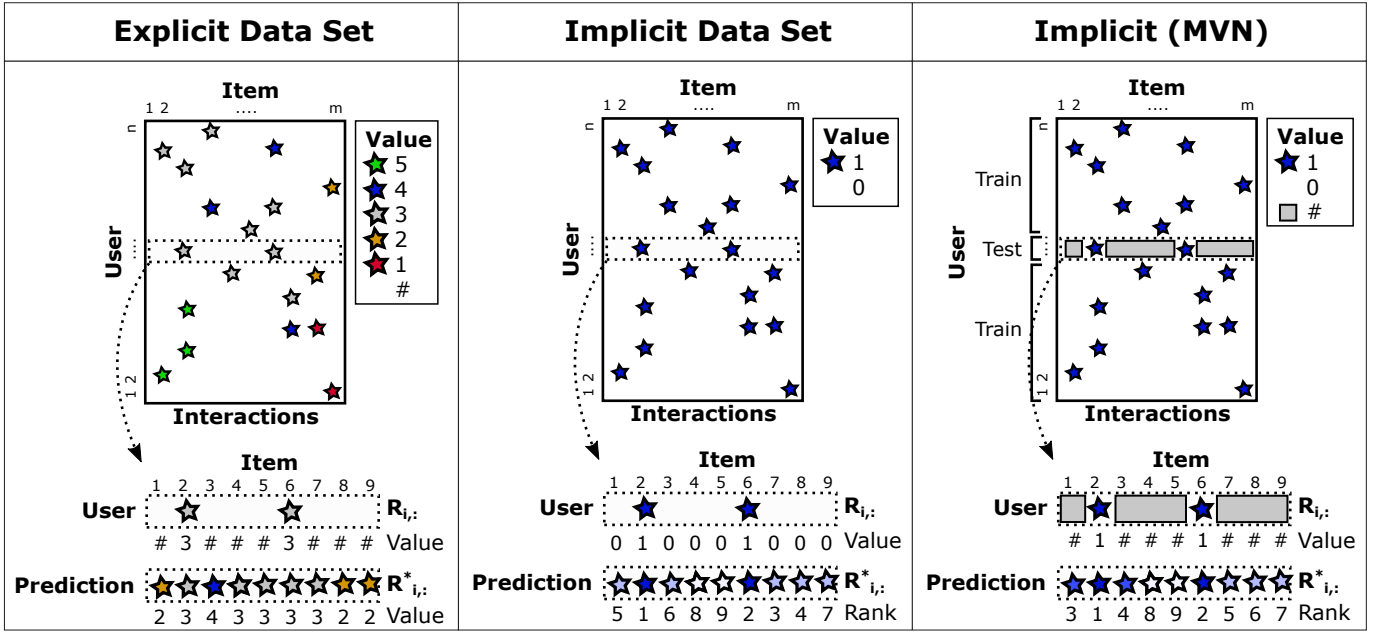}}
\caption{Illustration of an explicit data set of movie ratings (1-5) with missing ratings (\#) and the corresponding implicit data set of movie watching status (0/1) without missing watching status. The Multivariate Normal Distribution (MVN) motivates the reinterpretation of implicit data as training with a complete interaction matrix (0/1) and testing with prediction of the missing interactions (\#) based on known interactions (1).}
\label{figure:data}
\end{figure*}


In this study, we abstract the problem as predicting interactions of users and items. Suppose that we have $n$ users and $m$ items. Denote users $i\in\{1,2,...,n\}$ and items $j\in\{1,2,...,m\}$. The user and item interactions are stored in a $n \times m$ binary interaction matrix $\interactions\in\mathbb{R}^{n \times m}$. The binary entry $\interactions_{i,j}=\mathbb{I}(\text{user }i\text{ interacts with item }j)$ may denote whether user has watched a movie, played a video game, liked a joke, purchased a product, etc. In general the interactions do not have to be binary, because some data sets also record the interaction strength. This is a typical implicit recommendation task without missing entries, because the interactions or lack thereof are assumed to be known for every user. We say the matrix is complete, in contrast to incomplete matrices where some of the entries can be missing. This is illustrated in Figure~\ref{figure:data}, which also demonstrates the idea behind our model that we present next. For example, the user $i$ may have watched the first and the third movie:

\begin{equation}
\begin{array}{c}
\interactions_{i,:} = (1, 0, 1, 0, ..., 0)
\end{array}
\end{equation}

The task is then to predict the ranking of movies that the user has not yet watched but might like to watch. This is a prediction matrix $\predictions\in\mathbb{R}^{n \times m}$ where only the order of the values matters for ranking. For example, the model might predict values for all the $m$ movies as follows:

\begin{equation}
\begin{array}{c}
\predictions_{i,:} = (1.41, 0.10, 0.82, 0.04, ..., 0.21)
\end{array}
\end{equation}

The recommendation list for a user is obtained by taking movies that have the $k$ largest predicted values in $\predictions_{i,:}$, excluding those that the user has already watched. 

Since we are interested in recommending a list of items, we need to use ranking accuracy metrics to evaluate the quality of recommendations. These metrics compare the list of recommended items to the items the user is known to have interactions with. We used Precision@k and nDCG@k to measure different aspects of the recommendation accuracy. Precision@k is a useful measure of practical recommendation list accuracy when $k$ is typically small and the position of items do not matter. We chose $k=20$ to reflect a scenario of receiving Top-20 recommendations. We also used nDCG@m, meaning the nDCG evaluated over the entire recommendation list of $m$ items, to provide a wider perspective of the recommendation list accuracy. These metrics are defined as follows:

\subsubsection{Precision@k}
Assume for user $i$ the model produces a recommendation list $\reclist$, which is a permutation of items $\{1,2,...,m\}$ where the items that have interactions with the user in the training set have been excluded. The element $\reclist_{j}$ is the $j$'th item in the recommendation list, i.e. the index of $j$'th largest predicted value in $\predictions_{i,:}$. The Precision@k metric counts the number of items the user has interacted with, as a fraction of all items in the recommendation list of length $k$. The metric for the data set is the average over users:

\begin{equation}
\begin{array}{c}
\frac{1}{n}\sum_{i=1}^{n}\frac{1}{k}\sum_{j=1}^{k} \mathbb{I}(\text{user } i \text{ interacts with item } \reclist_{j})
\end{array}
\end{equation}



\subsubsection{nDCG@k}
The normalized Discounted Cumulative Gain at $k$ (nDCG@k) metric measures the position of interactions in the recommendation list. When a user interacts with an item, its position in the user's recommendation list is rewarded by the inverse of its logarithmic rank. These are called the discounted cumulative gains. Larger values for the metric are obtained by moving interactions towards the top of the list. In the optimal ranking, we have every true interaction on the top of the recommendation list, of the total of $m_i=|\{j : \interactions_{i,j}^{\text{(test)}} = 1\}|$, and the discounted cumulative gain has the value $\text{IDCG}_i=\sum_{j=1}^{\text{min}(m_i,k)}1/\text{log}_2(j+1)$. The nDCG@k is the discounted cumulative gain in the recommendation list $\reclist$ of length $k$, normalized by the maximum attainable value $\text{IDCG}_i$. The nDCG@k for the data set is the average over users:

\begin{equation}
\begin{array}{c}
\frac{1}{n}\sum_{i=1}^{n}\frac{1}{\text{IDCG}_i}\sum_{j=1}^{k} \frac{\mathbb{I}(\text{user } i \text{ interacts with item } \reclist_{j})}{\text{log}_2(j+1)}
\end{array}
\end{equation}

\section{Methods}

\subsection{New method: Multivariate Normal Distribution (MVN)}

First we present a simple new collaborative filtering model which has a competitive accuracy against baselines in implicit data sets. This simplicity implies a faster training time, more robustness due to fewer hyperparameter choices, and even an intuitive interpretation of the parameters. Suppose that every row of the interaction matrix is an observation from a multivariate normal distribution: $\interactions_{i,:} \sim\mathcal{N}(\mean,\covariance)$ with an unknown mean vector $\mean\in\mathbb{R}^{m}$ and an unknown covariance matrix $\covariance\in\mathbb{R}^{m \times m}$. Because the interaction matrix is complete, this model has a straightforward closed form solution for the distribution parameters. The maximum likelihood estimate is the sample mean vector and the sample covariance matrix:
\begin{equation}
\begin{array}{c}
\mean_{j} = \frac{1}{n}\sum_{s=1}^{n} \interactions_{s,j}
\end{array}
\end{equation}

\begin{equation}
\begin{array}{c}
\covariance_{i,j} = \frac{1}{n}\sum_{s=1}^{n}(\interactions_{s,i}-\mean_i)(\interactions_{s,j}-\mean_j)
\end{array}
\end{equation}

The crucial part is to interpret the data as follows. When we train the model, i.e. compute the mean vector and the covariance matrix, we assume that the data set is complete so that all user and item interaction values are either 'interaction' or 'no interaction'. However, when we predict for a given user we assume that the 'interaction' values are observed and the 'no interaction' values are missing.  This means that the user has a known interaction with these items, and we would like to predict the probability of interaction for every remaining item. This simple idea is illustrated in Figure~\ref{figure:data}.

This predicted probability has a straightforward closed form solution for the normal distribution. For a given user $i$, denote the indices of interactions as $\liked$ and the indices of no interactions $\notliked$ so that $\liked\cup\notliked=\{1,2,...m\}$. We use indexing $\interactions_{\notliked,\liked}$ to denote the submatrix with rows from $\notliked$ and columns from $\liked$. The predictions for the missing interactions are then given by the expectation of the conditional distribution $\interactions_{\{i\},\notliked}^{*} = \mathrm{E}(\interactions_{\{i\},\notliked}|\interactions_{\{i\},\liked})$. This can be shown to equal: 
\begin{equation}\label{eq:mvn}
\begin{array}{c}
\interactions_{\{i\},\notliked}^{*} = \mean_{\notliked}^{T} 
+ (\interactions_{\{i\},\liked}-\mean_{\liked})^{T}(\covariance_{\liked,\liked})^{-1}
\covariance_{\liked,\notliked}
\end{array}
\end{equation}

If the set of interactions is empty $\liked=\emptyset$ we define the model to predict the expected value $\interactions_{\{i\},:}^{*} = \mathrm{E}(\interactions_{\{i\},:}) = \mean$.

The standard MVN in previous equation considers the non-interactions $\notliked$ as missing data to be predicted. We could also consider the non-interactions as observed. In this method, we would predict the probability of user's interaction with item $j$ based on user's all other interactions and non-interactions $\other = \{1,...,m\} \setminus \{j\}$. The predictions are again given by the expectation of the conditional distribution $\predictions_{i,j} = \mathrm{E}(\interactions_{i,j}|\interactions_{i,\other})$. This equals: 
\begin{equation}\label{eq:mvn_obs}
\begin{array}{c}
\predictions_{i,j} = \mean_{j} 
+ (\interactions_{i,\other}-\mean_{\other})^{T}(\covariance_{\other,\other})^{-1}
\covariance_{\other,j}
\end{array}
\end{equation}

The modified MVN formula in Equation~\ref{eq:mvn_obs} predicts every interaction based on all other interactions and non-interactions. However, it is computationally challenging to apply directly because it requires the inversion of a large $(m-1) \times (m-1)$ matrix for every item. We can use the Sherman–Morrison formula as a computational shortcut so that only one matrix inversion is required \cite{shermanmorrison}. 

To regularize the algorithm, one option is to use the 'shrinkage' estimator of the covariance matrix \cite{shrinkage}. The coefficient $\alpha\in[0,1]$ determines the degree to which we interpolate between the observed covariance matrix and a covariance matrix of interactions being independent:
\begin{equation}\label{eq:shrinkage1}
\begin{array}{c}
\covariance_{\alpha} = (1-\alpha)\covariance + \alpha \frac{\text{Trace}(\covariance)}{m}\mathbf{I}
\end{array}
\end{equation}

A similar shrinkage estimator can be defined for the mean vector. The coefficient $\beta\in[0,1]$ determines the degree to which we interpolate between the observed item popularity and items being equally popular:
\begin{equation}\label{eq:shrinkage2}
\begin{array}{c}
\mean_{\beta} = (1-\beta)\mean + \beta \frac{\text{Sum}(\mean)}{m}\overline{1}
\end{array}
\end{equation}

Many methods that have been optimized for accuracy tend to recommend the most popular items and this 'popularity bias' has been found to affect many recommender systems \cite{topn0}. To predict without item popularity affecting the results at all, we can use Equation~(\ref{eq:mvn}) where we set the mean vector $\mean_{\notliked}:=0$ and substitute the sample covariance matrix with the sample correlation matrix $\covariance_{\liked,\liked}:=\correlation_{\liked,\liked}$ and $\covariance_{\liked,\notliked}:=\correlation_{\liked,\notliked}$. The correlation matrix is the normalized covariance matrix $\correlation_{i,j} = \frac{\covariance_{i,j}}{\sd_i\sd_j}$ where $\sd_{j}^{2} = \frac{1}{n}\sum_{i} (R_{i,j} - \mean_j)^{2}$ is the standard deviation. 
This is equivalent to taking the interaction matrix $\interactions$, then mean centering and normalizing by standard deviation column wise before the model is applied. 

\subsection{Baseline Methods}

\subsubsection{Random}

The random model is very simple: the recommendation list is a random permutation all items. The predicted recommendations can be sampled from a standard normal distribution for example:

\begin{equation}
\begin{array}{c}
\predictions_{i,j} \sim\mathcal{N}(0, 1)
\end{array}
\end{equation}

\subsubsection{Popularity}

We define item popularity as the fraction of users who interact with an item. Denoting $\mean_i=\sum_{i} \interactions_{i,j}/n$, the popularity based ranking predicts the same recommendation list of most popular items for every user:

\begin{equation}
\begin{array}{c}
\predictions_{i,j} = \mean_j
\end{array}
\end{equation}

\subsubsection{k Nearest Neighbour (kNN)}

The kNN is a straightforward recommendation method. In this method, we have defined a way to measure similarity between any two items. This results in an $m \times m$ item similarity matrix $\similarity\in\mathbb{R}^{m \times m}$. We define the similarity function as the Cosine similarity $\similarity_{i,j}=\frac{\sum_{s}\interactions_{s,i}\interactions_{s,j}}{\sqrt{\sum_{s}\interactions_{s,i}^2}\sqrt{\sum_{s}\interactions_{s,j}^2}}$. The rating prediction for item $j$ only considers the $k$ most similar items that have a value in the interaction matrix for user $i$. Denote this set of most similar items $\neighbourhood(i,j)$. Because user $i$ always has every rating (interaction or non-interaction), this neighbourhood for item $j$ is independent of the user $\neighbourhood(i,j)=\neighbourhood(j)$ and corresponds to the $k$ most similar items. The prediction for a user is then the similarity weighted average to the user's interaction status of $k$ most similar items: 

\begin{equation}
\label{similarity}
\begin{array}{c}
\predictions_{i,j} = \sum_{s\in \neighbourhood(i,j)} \similarity_{j,s} \interactions_{i,s}/\sum_{s\in \neighbourhood(i,j)} \similarity_{j,s}
\end{array}
\end{equation}

As others have pointed out \cite{topn0}, the normalizing denominator is not necessary for the ranking task and we in fact obtained much better predictions without it. This trick was utilized in some of the software packages and in the experiments we therefore predict simply by $\predictions_{i,j} = \sum_{s\in \neighbourhood(i,j)} \similarity_{j,s} \interactions_{i,s}$
. In this modified model we do not normalize the predictions to the interval $[0,1]$. However, this has further implications. We now show that dropping the normalization constant in Equation~(\ref{similarity}) implies that only the items with interactions are considered as observed and non-interacted items are missing. Let $\similarity^{(k)}$ denote the similarity matrix with only $k$ largest values in each column. The predictions can then be expressed:

\begin{equation}
\label{eq:knnpred}
\begin{array}{c}
\predictions = \interactions \similarity^{(k)}
\end{array}
\end{equation}

Because every row $\interactions_{i,:}$ is a binary vector with ones corresponding to items with interactions $\liked$, each predicted row $\predictions_{i,:}$ is:
\begin{equation}
\label{eq:knnpred2}
\begin{array}{c}
\predictions_{i,:} = \begin{pmatrix} \frac{1}{|\liked|}\sum_{s\in\liked\cap\neighbourhood(1)}\similarity_{s,1} & ... & \frac{1}{|\liked|}\sum_{s\in\liked\cap\neighbourhood(m)}\similarity_{s,m} \end{pmatrix}
\end{array}
\end{equation}

where we have divided by $|\liked|$ because multiplying by a constant does not change the ranking of items. For a maximum neighbourhood size $k=m$, the predictions are therefore the average similarity of items the user interacted with and the target item. For $k<m$ we do not count the similarity of an interacted item if it is not among $k$ most similar items. In this sense, the modified model bases the recommendations only on the items the user interacted with. The items user did not interact with are ranked based on their similarity to items the user interacted with. 

\subsubsection{Matrix Factorization (MF)}

The matrix factorization methods of dimension $d$ are defined in terms of $n \times d$ matrix $\users\in\mathbb{R}^{n \times d}$ of latent user factors as rows and $m \times d$ matrix $\items\in\mathbb{R}^{m \times d}$ of latent item factors as rows. The name matrix factorization comes from assuming that the interaction matrix is a product $\predictions=\users\items^T$ of these factor matrices. A prediction for user $i$ and item $j$ is therefore the product of the latent user vector $\users_{i,:}\in\mathbb{R}^d$ and the latent item vector $\items_{j,:}\in\mathbb{R}^d$:

\begin{equation}
\label{nobiases}
\begin{array}{c}
\predictions_{i,j} = \users_{i,:}\items_{j,:}^T
\end{array}
\end{equation}

Some implementations add user and item specific intercepts $\predictions_{i,j} = \users_{i,:}\items_{j,:}^T + \userbias_i + \itembias_j$, where $\userbias\in\mathbb{R}^n$ is a user bias vector and $\itembias\in\mathbf{R}^m$ is a item bias vector. The resulting model is in theory included in the above model. This can be seen by setting the first column of $\users_{i,:}$ to $1$ and the second column of $\items_{i,:}$ to $1$. Regularization can result in a some differences between the otherwise almost identical models. 

The latent vectors are initially unknown. One assumes that they are model parameters and finds them by fitting a regularized model to the data set of game likes: 
\begin{equation}
\label{eq:svd}
\begin{array}{c}
\users, \items = \text{argmin}_{\users,\items} L_(\users,\items) + \lambda (\|\users\|^2_F + \|\items\|^2_F)
\end{array}
\end{equation}
The matrix norm $\|\|^2_F$ denotes the Frobenius norm. The loss function $L_(\users,\items)$ determines which matrix factorization method we are using. 

\subsubsection{Least Squares (LS)}

The least squares loss is analogous to the Singular Value Decomposition (SVD), by which name it is sometimes called in the literature. If the regularization parameter is set to zero $\lambda = 0$ and there is no missing data, this method is identical to the standard SVD:
\begin{equation}
\begin{array}{c}
L_{\text{LS}}(\users,\items) = \sum_{i=1}^{n}\sum_{j=1}^{m}(\interactions_{i,j}-\predictions_{i,j})^2
\end{array}
\end{equation}

The regularized MF-LS ($\lambda > 0$) arrives at a factorization $\predictions = \users \items^{T}\in\mathbb{R}^{n \times m}$ by finding $\users\in\mathbb{R}^{n \times d},\items\in\mathbb{R}^{m \times d}$ that minimize the least squares loss where the matrix norm $\|.\|^2_F$ denotes the Frobenius norm: 

\begin{equation}
\begin{array}{c}
\users, \items = \text{argmin}_{\users,\items} \|\interactions-\users \items^T\|^2_F + \lambda \|\users\|^2_F + \lambda \|\items\|^2_F
\end{array}
\end{equation}


An iterative approach is required to solve this problem. In the Alternating Least Squares (ALS) approach \cite{topn2}, either the latent item vectors $\items$ or the latent user vectors $\users$ are assumed to be fixed and the optimal solution for the other is found. The optimization starts by initializing $\users$ and $\items$ with random values. We can then solve with standard ridge regression either 

\begin{equation}
\begin{array}{c}
\users^T = (\items^T \items + \lambda \mathbf{I})^{-1} \items^T \interactions^T
\end{array}
\end{equation}

\begin{equation}
\begin{array}{c}
\items = (\users^T \users + \lambda \mathbf{I})^{-1} \users^T \interactions
\end{array}
\end{equation}

We iterate between fixing latent item factors $\items$ to find optimal values for latent user factors $\users$, or fixing the resulting latent user factors $\users$ to find optimal values for latent item factors $\items$. The iterations are repeated until convergence. This is the standard MF-LS algorithm.

We can modify the method to regard non-interactions as missing as follows. First we run the standard MF-LS model to convergence to find a factorization $\predictions = \users \items^{T}$. Then, to predict for user $i$ we learn a new latent user vector $\usersmissing_{i,:}\in\mathbb{R}^{1 \times m}$ based on only the items $\liked$ the user has interactions with, considering the non-interactions as 'missing' data: 

\begin{equation}
\begin{array}{c}
\usersmissing_{i,:}^{T} = (\items_{\liked,:}^{T} \items_{\liked,:} + \lambda \mathbf{I})^{-1} \items_{\liked,:}^{T} \interactions_{i,\liked}^{T}
\end{array}
\end{equation}

The predictions $\predictions_{i,:} $ for a user can then be given by a model that considers the non-interactions as observed or missing:

\begin{equation}
\begin{array}{ll}
\predictions_{i,:} = \users_{i,:} \items^{T} & (\text{observed}) \\
\predictions_{i,:} = \usersmissing_{i,:} \items^{T} & (\text{missing}) \\
\end{array}
\end{equation}

\subsubsection{Bayesian Personalized Ranking (BPR)}

The BPR loss is based on a probabilistic model, where the predictions are defined as to maximize the likelihood of ranking an interaction higher than no interaction. For user $i$, define the set of all pairs where one item was interacted with and the other was not as $\mathcal{P}^{(i)}=\{(j,s) | \interactions_{i,j}=1, \interactions_{i,s}=0\}$. The probability of interacting with one item and not interacting with the other is then defined as $P_{i,j,s}=1/(1+\text{exp}(-(\predictions_{i,j}-\predictions_{i,s})))$. The loss is defined as the negative log-likelihood of the probabilistic model: 

\begin{equation}
\begin{array}{c}
L_{\text{BPR}}(\users,\items) = -\sum_{i=1}^{n}\sum_{j,s\in \mathcal{P}^{(i)}} \text{log}[P_{i,j,s}]
\end{array}
\end{equation}

\subsubsection{Logistic Matrix Factorization (LOG)}

The logistic loss is also based on a probabilistic model, which directly models the probability of interacting or not interacting with an item in the interaction matrix. The probability of user interacting with an item is defined as $P_{i,j}=1/(1+\text{exp}(-\predictions_{i,j}))$. The loss is defined as the negative log-likelihood of observed interactions:
\begin{equation}
\begin{array}{c}
L_{\text{LOG}}(\users,\items) = -\sum_{i=1}^{n}\sum_{j=1}^{m} \text{log}[P_{i,j}^{\interactions_{i,j}}(1-P_{i,j})^{1-\interactions_{i,j}}]
\end{array}
\end{equation}




\subsection{MVN, SVD and KNN as regression}

In this section, we derive mathematical results that explain the differences between the MVN, kNN and SVD methods. We prove that these can be interpreted as regression models. The main difference is that the MVN regards the non-interactions as missing data to be be predicted. We showed previously that the standard SVD and kNN baselines can also be modified to consider them as missing. In the experiments we find that the models where non-interactions do not influence predictions perform much better and do not require regularization of latent dimensions or number of neighbours. 

\subsubsection{MVN}

In the following, for simplicity of notation we assume a pre-centered data matrix $\centered$ i.e. $\centered = \interactions - 1 \mean^{T}$ where $\interactions\in\mathbb{R}^{n \times m}$ is the user, item interaction matrix.  For example, $\interactions$ can denote the binary like status of every user and movie pair. 

\begin{theorem}
For a given user $i$, the MVN model is equivalent to least squares regression $\centered_{:,\liked} \coefmat = \centered_{:,\notliked}$ where we use the items user interacted with $\liked\subseteq\{1,2,…,m\}$ to predict the probability of interaction with the remaining items $\notliked=\{1,2,…,m\}\setminus\liked$.
\end{theorem}
Proof: To predict the probability of interaction with item $j$ from the interaction status of items $\liked$, we can make a least squares linear regression to predict the value of column $j$ from the values of columns $\liked$ where $\coefvec\in\mathbb{R}^{|\liked|}$: 
\[
\centered_{:,\liked} \coefvec=\centered_{:,j} \Longrightarrow \coefvec = (\centered_{:,\liked}^T \centered_{:,\liked})^{-1} \centered_{:,\liked}^T \centered_{:,j}
\]
This generalizes to predicting the probability of multiple interactions $\notliked$ simultaneously where $\coefmat\in\mathbb{R}^{|\liked|\times|\notliked|}$:
\[
\centered_{:,\liked} \coefmat=\centered_{:,\notliked} \Longrightarrow \coefmat = (\centered_{:,\liked}^T \centered_{:,\liked})^{-1} \centered_{:,\liked}^T \centered_{:,\notliked}
\]
Given a given user $i$ with known interactions $\centered_{i,\liked}$ the remaining interactions $\centered_{i,\notliked}^{*}$ can be predicted:
\[
\centered_{i,\notliked}^{*} = \centered_{i,\liked} \coefmat 
\]
Denote the following factorization of the covariance matrix $\covariance=\frac{1}{n} \centered^T \centered$ where $\covariance\in\mathbb{R}^{m \times m}$ corresponding to columns $\liked$ and $\notliked$ of the data matrix:
\[
\covariance = 
\begin{pmatrix}
\covariance_{\liked,\liked} & \covariance_{\liked,\notliked} \\
\covariance_{\notliked,\liked} & \covariance_{\notliked,\notliked} 
\end{pmatrix}
\]
In fact, we directly have $\coefmat = (\centered_{:,\liked}^T \centered_{:,\liked})^{-1} \centered_{:,\liked}^T \centered_{:,\notliked} = \covariance_{\liked,\liked}^{-1}\covariance_{\liked,\notliked}$ and therefore 
\[
\centered_{i,\notliked}^{*} = \centered_{i,\liked} \coefmat  
\Longrightarrow 
(\predictions_{i,\notliked} - \mean_{\notliked}^{T}) = (\interactions_{i,\liked} - \mean_{\liked}^{T}) \covariance_{\liked,\liked}^{-1}\covariance_{\liked,\notliked}
\]
This equals Equation~(\ref{eq:mvn}) for the MVN model $\square$.

The shrinkage estimator in Equation~(\ref{eq:shrinkage1}) is equivalent to adding Tikhonov (l2) regularization penalty to the least squares problem for some value of $\lambda$:
\[
\text{argmin}_{B} =
 || \centered_{:,\liked} \coefmat - \centered_{:,\notliked} ||^2 + \lambda || \coefmat ||^2 
\]
\[\Longrightarrow \coefmat = (\centered_{:,\liked}^T \centered_{:,\liked} + \lambda \mathbf{I})^{-1} \centered_{:,\liked}^T \centered_{:,\notliked}\]


\subsubsection{SVD}

We consider the MF-LS model without regularization because this equals the standard SVD when the implicit interaction matrix $\centered$ has no missing entries.

\begin{theorem}
For a given user $i$, the MF-LS model ($d$ latent factors and $\lambda = 0$) is equivalent to total least squares regression $\centered_{:,\reference} \coefmat = \centered_{:,\remaining}$  where we use arbitrary items $\reference\subseteq\{1,2,…,m\}$ with $|\reference|=d$ to predict the probability of interaction with other items $\remaining=\{1,2,…,m\}\setminus\reference$. The formulation is symmetric in the sense that it does not matter which items are chosen as the reference items $\reference$.
\end{theorem}
Proof: The proof is a corollary of the following two results: 

\begin{enumerate}
    \item Eckart-Young theorem: Least squares low-rank approximation of $\centered$ of rank $d$ $\Longleftrightarrow$ Singular Value Decomposition of $\centered$ with $d$ largest singular values
    \item Singular Value Decomposition of $\centered$ with $d$ largest singular values $\Longleftrightarrow$ Total least squares regression of data matrix $\centered$ with the formulation $\centered_{:,\reference} \coefmat=\centered_{:,\remaining}$ where $|\reference|=d$ is any set of columns.
    \cite{tlsoverview}
\end{enumerate}

Denote the Singular Value Decomposition (SVD) of data:
\[\centered = \leftsing \valsing \rightsing^T\]
where $\valsing=\text{diag}(\sign_1, ..., \sign_m)$ is the diagonal matrix of decreasing singular values, $\leftsing$ and $\rightsing$ are both orthogonal matrices with left and right singular vectors as columns. 

Assume we have fitted the MF-LS model to convergence with $d$ latent factors and no regularization. By the Eckart-Young theorem this is equivalent to the clipped SVD: 
\[\centered^{*} = \leftsing \valsing_d \rightsing^T\]
where $\valsing_d=\text{diag}(\sign_1, ..., \sign_d,0,...,0)$. For a given user $i$, the predictions are simply the $i$'th row  $\centered^{*}_{i,:}$. By the second result, these predictions are equivalent to total least squares regression of data matrix $\centered$ with the formulation $\centered_{:,\reference} \coefmat=\centered_{:,\remaining}$ where $|\reference|=d$ is any set of columns. $\square$

\begin{theorem}
For a given user $i$, the MF-LS model ($d$ latent factors and $\lambda = 0$) is also equivalent to least squares regression $\centered \coefmat = \centered$  where we predict all interactions from all interactions and non-interactions, including interactions with the item itself, but regularize the problem by taking the pseudo-inverse $\centered^\dagger$ with only the $d$ largest singular values. 
\end{theorem}
Proof: Using the standard SVD, the pseudo-inverse of $\centered$ with only the $d$ largest non-zero singular values can be defined:
\[\centered^{\dagger}_{d} = \rightsing \valsing^{-1}_{d} \leftsing^T\]
where $\valsing^{-1}_{d}=\text{diag}(1/\sign_1, ..., 1/\sign_d,0,...,0)$. 
By the theorem statement we calculate $\centered \coefmat = \centered \Longrightarrow \coefmat = \centered^{\dagger}_{d}\centered$ and therefore
\[\centered^{*} = \centered \coefmat = \centered \centered^{\dagger}_{d}\centered = \leftsing \valsing_{d} \rightsing^T\]

which by the previous results is equal to the clipped SVD, or the best least squares low-rank approximation of rank $d$. $\square$

The SVD has a connection to the MVN, since the covariance matrix has eigenvectors $\rightsing_{:,j}$ and eigenvalues $\sign^2_j/n$:
\[
\covariance = \frac{1}{n} \centered^T \centered = \frac{1}{n} \rightsing \valsing^2 \rightsing^T
\]

Even though the regularization parameter was set to $\lambda=0$, this model regularizes by limiting the number of latent factors $d$. One way to understand this is to define the projection matrix $\projection = \rightsing \mathbf{I}_d \rightsing^T$ where $\mathbf{I}_d=\text{diag}(1, ..., 1,0,...,0)$ with ones up to the d'th value. This projects a vector to the $d$ largest right singular vectors. Now we have

\[\centered \projection =  \leftsing \valsing \rightsing^T \rightsing \mathbf{1}_d \rightsing^T = \leftsing \valsing_d \rightsing^T = \centered^{*} \]

For user $i$ with observations $\centered_{i,:}$ the predictions $\centered_{i,:}^{*}$ are obtained by projection $\centered_{i,:}^{*} = \centered_{i,:} \projection$ into the eigenvectors corresponding to the $d$ largest eigenvalues of the covariance matrix. The eigenvectors form a complete orthonormal basis and taking all of the latent factors $d=m$ would simply predict the observed values: $\centered \projection= \leftsing \valsing \rightsing^T \rightsing \mathbf{I}_m \rightsing^T = \leftsing \valsing \rightsing^T = \centered$. Therefore in the SVD for complete data, we assume that we have observed all interactions and non-interactions of an user, and some have an error that is fixed by a projection into a lower-dimensional subspace of the covariance matrix. 

\subsubsection{kNN}
The standard kNN formulation in recommender systems can be interpreted as kernelized regression.

\begin{theorem}
For a given user $i$, the kNN model ($k=m$) is equivalent to Nadaraya–Watson kernel regression $\interactions_{i,:}=f(\interactions)$ where we use all items $\{1,2,…,m\}$ to predict the probability of interaction with each item $\predictions_{i,j} = \mathrm{E}(\interactions_{i,j}|\interactions_{:,j})$ based on the kernel similarity of items $\interactions_{:,j}$ where the column of user interactions are the features for that item.

\end{theorem}

Proof: The Nadaraya–Watson regression is formulated as

\begin{equation}
\begin{array}{c}
\label{eq:nwregression}
\mathrm{E}(y|x) = \frac{\sum_{j=1}^{m}K(x,x_j)y_j}{\sum_{j=1}^{m}K(x,x_j)}
\end{array}
\end{equation}
where $K(.)$ is a kernel function and $\{(x_j,y_j)\}_{j=1}^{m}$ is the sample of item and outcome pairs. For every user $i$ we can define a data set $x_j = \interactions_{:,j}\in\mathbb{R}^n$ and $y_j = \interactions_{i,j}\in\mathbb{R}$. This corresponds to predicting the probability of interaction with every item based on the corresponding column of $\interactions$, where each item is an input and the column contains the user interactions as features. We define the kernel as the cosine similarity $\similarity_{j,k} = K(\interactions_{:,j},\interactions_{:,k}) =  \frac{\sum_{s=1}^{n}\interactions_{s,j}\interactions_{s,k}}{\sqrt{\sum_{s=1}^{n}\interactions_{s,j}^2}\sqrt{\sum_{s=1}^{n}\interactions_{s,k}^2}}$, which we can interpret as a density kernel in a latent high dimensional feature space. The Equation~(\ref{eq:nwregression}) then becomes:

\begin{equation}
\begin{array}{c}
\label{eq:sim}
\predictions_{i,j} = \mathrm{E}(\interactions_{i,j}|\interactions_{:,j})
 = \frac{\sum_{k=1}^{m}\similarity_{j,k}\interactions_{i,k}}{\sum_{k=1}^{m}\similarity_{j,k}}
\end{array}
\end{equation}

Which is equal to the kNN model in Equation~(\ref{similarity}) where $\neighbourhood(i,j)=\{1,...,m\}$ for neighbourhood size $k=m$ $\square$.

To adjust regularization it is possible to consider neighbourhood sizes $k$ smaller than $m$, where take the interaction status into account only if the item belongs to $k$ most similar item to the target item.

\section{Results}

\subsection{Data sets and processing}

\begin{table}[htbp]
\caption{Statistics of Each Data Set}
\begin{center}
\begin{tabular}{|r|r|r|r|r|}
\hline
\textbf{Data set} & \textbf{\textit{Users}} & \textbf{\textit{Items}} & \textbf{\textit{Interactions}}& \textbf{\textit{Processing}}\\
\hline
ml-100k & 943 & 1682 & 100000 & \\
\hline
ml-1m & 6040 & 3706 & 1000209 & \\
\hline
ml-10m & 69878 & 10677 & 10000054 & \\
\hline
jester-1 & 24573 & 100 & 713090  & $\mathbb{I}(\interactions_{i,j}>3)$\\
\hline
jester-2 & 23004 & 100 & 657954 & $\mathbb{I}(\interactions_{i,j}>3)$\\
\hline
jester-3 & 23167 & 99 & 222516 & $\mathbb{I}(\interactions_{i,j}>3)$\\
\hline
movielens-2k-v2 & 2113 & 10109 & 855598 & \\
\hline
lastfm-2k & 1892 & 17632 & 92834 & \\
\hline
delicious-2k & 1867 & 69223 & 104799 & \\
\hline
book-crossing & 13417  & 14504  & 156632  & $\mathbb{I}(\interactions_{i,j}>0)$\\
\hline
book-implicit & 12724 & 27805 & 353448  & $\mathbb{I}(\interactions_{i,j}=0)$\\
\hline
\end{tabular}
\label{tab1}
\end{center}
\label{table:stats}
\end{table}

We used public data sets displayed in Table~\ref{table:stats}. The Movie Lens (ml-100k, ml-1m, ml-10m) data sets consists of increasing number of user and movie ratings \cite{movielens}. The Jester (jester-1, jester-2, jester-3) are closely related joke rating data sets \cite{jester}. The Hetrec data sets (movielens-2k-v2, lastfm-2k, delicious-2k) were published together in a conference with about 2000 users in each \cite{hetrec}. The book crossing (book-crossing, book-implicit) data sets consist of explicit and implicit user and book interactions \cite{bookcrossing}. Most of the data sets were originally gathered and interpreted as explicit data, but in this study we transform them into implicit data.

We define each user and item pair as a potential interaction. The processing column indicates which pairs where considered as an interaction, no processing indicates that every pair in the data set was considered as an interaction. For example, in the Movie Lens rating data set we assume that users have watched the movies they rated and haven't yet watch the movies they didn't rate. The task is then to recommend interesting new movies the user would like to watch based on this data. In the Jester data a rating greater than 3 was regarded as 'liking' the joke and other ratings and missing joke ratings were considered as 'not liking' a joke. The task is to recommend which of the not yet liked jokes the user would like. The 'book crossing' data set actually contains two data sets, one based on explicit ratings $\interactions_{i,j} \neq 0$ and another based on implicit interactions $\interactions_{i,j} = 0$. We therefore used this data set as two separate data sets 'book-crossing' and 'book-implicit'. 

Many of the data sets have already been filtered for users or items by requiring minimum number of interactions per user or item. We did not do additional filtering, but we had to limit the number of users and items in the book crossing data set because the baseline methods would take a very long time or run out of memory with the original 105 283 users and 340 556 items. We therefore took only items with at least 5 users, and then only those users with at least 3 items. The resulting user and item counts are displayed in the Table~\ref{table:stats}.


\subsection{Model implementation}

Because the software library may have a significant effect on the results, we trained models reported in the experiments with different libraries to verify that the results are consistent. We used the Python matrix algebra library 'numpy' to make straightforward baseline implementations (Random, Popularity, MVN, kNN, MF-LS). We used the Python recommender system library 'implicit' (kNN, MF-LS, MF-BPR, MF-LOG) and the Python factorization machine library 'lightfm' (MF-BPR, MF-LOG). These libraries are designed for implicit data sets, but they are not as popular as libraries for explicit data. We therefore verified that better results were not obtained with the fastfm (MF-LS, MF-LOG, MF-BPR) and libfm (MF-LS, MF-LOG) libraries. 
We obtained best and identical baseline results with the implicit library and our own numpy implementations, lightfm also achieved good results. The exception is the MF-LOG model which seemed sub-optimal in both implicit and lightfm. However, when we implemented this model with fastfm or libfm it still did not reach the performance of other models. 


\subsection{Evaluation by cross validation}

We use 5-fold cross validation to evaluate the methods. In explicit data sets, the cross-validation is often implemented by a straightforward random split of 20\% of interactions to the test fold and 80\% of the interactions to the training fold. The RMSE or other metric is then calculated by comparing the true values and predicted values on the 20\% of test interactions. For implicit data sets, we need to compare how well the methods predict every observed 'interaction' over all remaining 'non-interaction' values for users in the test set. 

We therefore define the evaluation as follows: in 5-fold cross validation we randomly sample 20\% of users to the test fold, and randomly take their $s$ interactions as a seed that belongs to the train fold, where their remaining interactions are assigned to the test fold. The train fold therefore contains 80\% of users and 20\% of users with only $s$ interactions. The test fold contains the 20\% of users with interactions that are not in the training set, i.e. those interactions that remain after removing the $s$ seed interactions for each test user. We later investigate what effect the choice of $s$ has on the results. In the other experiments we used $s=3$. This simulates the setting of a user mentioning 3 favourite movies who then receives an ordered list of all other movies. 

\subsection{Hyperparameter selection}

The models have hyperparameters that need to be tuned to obtain optimal performance. Because training a model with an extensive hyperparameter selection grid can be very time-consuming, some studies have included a model with few different hyperparameter choices as separate models. However, there is no guarantee that these choices are optimal. Other authors have pointed out that newly proposed methods may not in fact increase the performance as long as the baseline matrix factorization method is very carefully set up \cite{compare1}. However, the amount of collective effort that occurred in the Netflix prize may be unfeasible in many applications, so those baselines that achieve good results with fewer hyperparameters or even default values are particularly attractive.

\begin{figure}
\centerline{\includegraphics[width=\columnwidth]{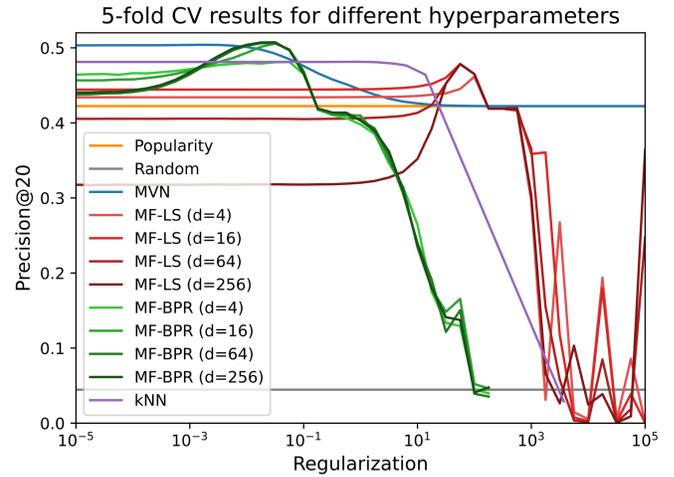}}
\caption{Different models have different sensitivity to hyperparameter choices, and some models (MF-LS, MF-BPR) require very careful choices to obtain their optimal performance in the Movie Lens 1M data set.}
\label{figure:hyperparameters}
\end{figure}

Figure~\ref{figure:hyperparameters} illustrates every possible regularization value $\lambda$ with $d=4,16,64,256$ latent factors in the Movie Lens 1M data set, which was the only data set where a baseline method (MF-BPR) achieved a slightly better accuracy. The regularization of MF-LS and MF-BPR models needs to be set very carefully, especially when the number of latent factors increases. The MVN and kNN are very attractive models from a practical point of view because they have only one hyperparameter that works for a wide range of regularization values and produces good results when set to zero. 
With increasing regularization the result converge either to the Popularity or Random baselines.

To be fair to the baseline models, we chose the hyperparameters carefully. In each training fold, we split 20\% of the training set users into a validation set. Our hyperparameter selection was based on optimizing the Precision@20 metric in the validation set. After we found the hyperparameters that maximize the metric the validation set, we re-trained the model with these choices on the whole training set. For each method, we chose an exponential grid of values where the end points were determined such that a clear concave maximum was found between them. There are no hyperparameters in the Random and Popularity models. The kNN has as a single neighbourhood size hyperparameter, which we evaluated on $k=1,2,4,8,...,m$. The MVN model has as a single optional regularization hyperparameter $\lambda\geq0$, which we evaluated on $\lambda\in\{10^{-5.00}, 10^{-4.75}, ..., 10^{4.75}, 10^{5.00}\}$. We evaluated the matrix factorization models on a grid with the number of latent factors $d=256$ and the same regularization parameters $\lambda\in\{10^{-5.00}, 10^{-4.75}, ..., 10^{4.75}, 10^{5.00}\}$. They were run for $100$ iterations over the data set, known as epochs. 

We note that in principle the matrix factorization models are even more complicated: they can be regularized by limiting the number of latent factors $d$, setting the regularization parameter $\lambda$, or stopping the iterations early. We obtained better results by setting the regularization parameter carefully and running to convergence rather than setting the regularization parameter to a small constant and stopping the iterations early. The number of latent factors could often be a large constant if the regularization parameter was chosen very carefully. The matrix factorization methods can have even more hyperparameters than presented here: different regularization for users/items or biases/interactions, iterations, learning rate, optimization method, etc. We believe that the comparison was fair, and in any case there are limitations in practise to how much effort can be taken to tweak a method. 

\subsection{Prediction of interactions}

\begin{table*}
\caption{5-fold CV: Precision@20 and nDCG@m in Different Data Sets for Default Hyperparameters ($d=64$, $\lambda=0$)}
\label{default}
\centering
\begin{tabularx}{\textwidth}{llllllllllll}
\textbf{Precision@20}       & ml-100k        & ml-1m          & ml-10m         & jester-1  & jester-2  & jester-3  & book-1  & book-2 & ml-2k-v2 & lastfm-2k      & delicious-2k   \\
Random     & 0.061          & 0.045          & 0.013          & 0.281          & 0.278          & 0.086          & 0.001                  & 0.001           & 0.040          & 0.003          & 0.001          \\
Popularity & 0.447          & 0.422          & 0.393          & 0.527          & 0.512          & 0.298          & 0.016                  & 0.028           & 0.674          & 0.200          & 0.009          \\
MVN        & \textbf{0.565} & \textbf{0.503} & \textbf{0.513} & \textbf{0.533} & \textbf{0.519} & \textbf{0.304} & \textbf{0.026}         & \textbf{0.029}  & 0.674          & \textbf{0.305} & \textbf{0.240} \\
kNN        & 0.544          & 0.481          & 0.489          & 0.528          & 0.514          & 0.303          & 0.017                  & 0.015           & 0.625          & 0.259          & NaN            \\
MF-LS      & 0.508          & 0.448          & 0.458          & 0.442          & 0.436          & 0.224          & 0.019                  & 0.026           & 0.566          & 0.299          & 0.006          \\
MF-BPR     & 0.551          & 0.452          & 0.371          & 0.392          & 0.394          & 0.091          & 0.008                  & 0.010           & \textbf{0.684} & 0.139          & 0.003          \\
MF-LOG     & 0.321          & 0.250          & 0.224          & 0.323          & 0.320          & 0.080          & 0.008                  & 0.013           & 0.319          & 0.156          & 0.003         \\
\textbf{nDCG@m} & ml-100k        & ml-1m          & ml-10m         & jester-1  & jester-2  & jester-3  & book-1  & book-2 & ml-2k-v2 & lastfm-2k      & delicious-2k   \\
Random     & 0.486          & 0.474          & 0.385          & 0.625          & 0.621          & 0.410          & 0.154                  & 0.172           & 0.518          & 0.303          & 0.267          \\
Popularity & 0.711          & 0.682          & 0.661          & 0.784          & 0.774          & 0.683          & 0.201                  & 0.228           & 0.788          & 0.518          & 0.267          \\
MVN        & \textbf{0.779} & \textbf{0.722} & \textbf{0.728} & \textbf{0.789} & \textbf{0.781} & \textbf{0.689} & \textbf{0.224}         & \textbf{0.229}  & \textbf{0.791} & 0.566          & \textbf{0.461} \\
kNN        & 0.771          & 0.720          & 0.726          & 0.787          & 0.779          & 0.688          & 0.206                  & 0.212           & 0.786          & 0.553          & NaN            \\
MF-LS      & 0.751          & 0.694          & 0.695          & 0.738          & 0.733          & 0.587          & 0.204                  & 0.218           & 0.740          & \textbf{0.571} & 0.274          \\
MF-BPR     & 0.767          & 0.691          & 0.653          & 0.712          & 0.710          & 0.449          & 0.176                  & 0.199           & 0.775          & 0.440          & 0.272          \\
MF-LOG     & 0.669          & 0.633          & 0.599          & 0.648          & 0.642          & 0.404          & 0.183                  & 0.200           & 0.695          & 0.432          & 0.299         
\end{tabularx}
\end{table*}

\begin{table*}
\caption{5-fold CV: Precision@20 and nDCG@m in Different Data Sets for Hyperparameters found in Validation Set}
\label{hyperparameters}
\centering
\begin{tabularx}{\textwidth}{llllllllllll}

\textbf{Precision@20}       & ml-100k        & ml-1m          & ml-10m         & jester-1  & jester-2  & jester-3  & book-1  & book-2 & ml-2k-v2 & lastfm-2k      & delicious-2k   \\
Random     & 0.061          & 0.045          & 0.013          & 0.281          & 0.278          & 0.086          & 0.001          & 0.001                  & 0.040           & 0.003          & 0.001          \\
Popularity & 0.447          & 0.422          & 0.393          & 0.527          & 0.512          & 0.298          & 0.016          & 0.028                  & 0.674           & 0.200          & 0.009          \\
MVN        & \textbf{0.569} & 0.503          & \textbf{0.513} & \textbf{0.533} & \textbf{0.519} & \textbf{0.304} & \textbf{0.032} & \textbf{0.034}         & 0.688           & \textbf{0.363} & \textbf{0.209} \\
kNN        & 0.545          & 0.481          & 0.490          & 0.528          & 0.514          & 0.303          & 0.019          & 0.015                  & 0.626           & 0.259          & NaN            \\
MF-LS      & 0.544          & 0.478          & 0.477          & 0.527          & 0.512          & 0.299          & 0.025          & 0.029                  & 0.669           & 0.312          & 0.022          \\
MF-BPR     & 0.567          & \textbf{0.507} & 0.484          & 0.524          & 0.508          & 0.276          & 0.007          & 0.017                  & \textbf{0.690}  & 0.162          & 0.001          \\
MF-LOG     & 0.390          & 0.427          & 0.393          & 0.527          & 0.512          & 0.298          & 0.012          & 0.018                  & 0.674           & 0.219          & 0.003         \\
\textbf{ncdg@m}           & ml-100k        & ml-1m          & ml-10m         & jester-1  & jester-2  & jester-3  & book-1  & book-2 & ml-2k-v2 & lastfm-2k      & delicious-2k   \\
Random     & 0.486          & 0.474          & 0.385          & 0.625          & 0.621          & 0.410          & 0.154                  & 0.172           & 0.518          & 0.303          & 0.267          \\
Popularity & 0.711          & 0.682          & 0.661          & 0.784          & 0.774          & 0.683          & 0.201                  & 0.228           & 0.788          & 0.518          & 0.267          \\
MVN        & \textbf{0.782} & \textbf{0.722} & \textbf{0.728} & \textbf{0.789} & \textbf{0.781} & \textbf{0.689} & \textbf{0.239}         & \textbf{0.238}  & \textbf{0.794} & \textbf{0.616} & \textbf{0.441} \\
kNN        & 0.769          & 0.720          & 0.724          & 0.787          & 0.779          & 0.687          & 0.195                  & 0.190           & 0.786          & 0.548          & NaN            \\
MF-LS      & 0.775          & 0.716          & 0.714          & 0.782          & 0.774          & 0.682          & 0.216                  & 0.226           & 0.784          & 0.582          & 0.339          \\
MF-BPR     & 0.776          & 0.718          & 0.702          & 0.779          & 0.767          & 0.666          & 0.177                  & 0.205           & 0.769          & 0.454          & 0.263          \\
MF-LOG     & 0.673          & 0.685          & 0.661          & 0.784          & 0.773          & 0.681          & 0.179                  & 0.204           & 0.754          & 0.455          & 0.303         
\end{tabularx}
\end{table*}

Our main experiment measures the ranking accuracy of the models in different data sets. We present the results with default hyperparameter choices in Table~\ref{default}, and the results with careful hyperparameter choices based on the validation set in Table~\ref{hyperparameters}. We used the Precision@20 and nDCG@m metrics as discussed previously. The Random, Popularity and MVN models are implemented with numpy and the kNN, MF-LS, MF-BPR and MF-LOG are based on the Implicit library. Similar results were obtained with the LightFM library. 

When we compare the two tables, we see that the results are already close to optimal for the MVN model with no regularization and the kNN model with the maximum number of neighbours. The results for kNN are missing in the delicious-2k data set because the baseline ran out of memory (64GiB) when it attempted to compute the similarity matrix. However, the matrix factorization models with default hyperparameters are far from optimal, and extensive search effort is required to achieve good results. Training models with a dense hyperparameter selection grid was found to be very time consuming. 
Therefore, it would be desirable for a model to achieve close to optimal results with a wider range of hyperparameter values and have fewer hyperparameters. 

Predicting the most popular items already has quite a lot of signal when compared to the random baseline. The popularity baseline is surprisingly good in the jester and movielens-2k-v2 data sets. This indicates that users tend have the same opinion which jokes are good, and they all tend to watch the most popular movies. The MVN is the best model in most data sets and reaches almost equal performance when it happens to be the second. Of the two most popular baselines, sometimes the kNN model is better (movie lens, jester) and sometimes the MF-LS is better (book crossing, lastfm). The ranking based MF-BPR model can sometimes achieve better results (movie lens), but it is not very robust because sometimes the performance is far from the best (book crossing, lastfm, delicious). The MF-LOG model appears to converge to the same results as predicting the most popular items. Based on these results, we think that the MVN and kNN model are the most useful in practise with the MVN obtaining better results by a small to significant margin depending on the data set.

\subsection{Runtime}

\begin{figure}
\centerline{\includegraphics[width=\columnwidth]{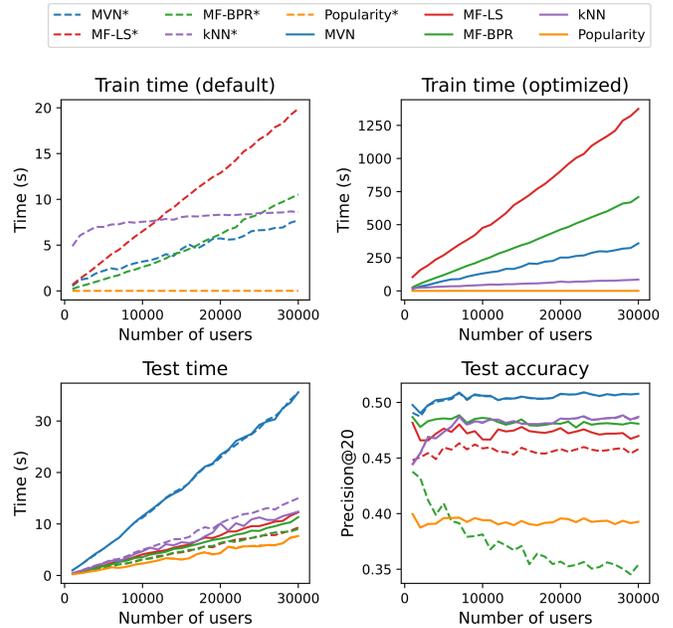}}
\caption{Runtime can be an important consideration. Models with default hyperparameters (dashed) have a similar cost for one combined training and test run over the data set, but optimizing for hyperparameters (solid) takes two orders of magnitude longer.}
\label{figure:runtime}
\end{figure}

Some models take extensive computational effort to train. We therefore varied the number of users in the Movie Lens 10M data set to have a simple estimate of the runtime cost. The full data set has 69 878 users, of whom we took a random sample of size $n=1000, 2000, ..., 30 000$ into both training and test set. The models were trained with the train users and seed interactions of test users, then tasked to predict the missing interactions of test users. We measured training time, test time, and resulting Precision@20. We considered training  models with both default hyperparameter choices and the extensive hyperparameter search grid.

Results are displayed in Figure~\ref{figure:runtime}. For models with default hyperparameters, one combined training and test run over all users takes a similar time for the models. Both runtimes go up linearly with the number of users. The MVN model is the fastest to train because it simply computes the covariance matrix, but slowest to test because it has an additional constant factor caused by the matrix inversion of the $k$ seed interactions. 
However, if it is necessary to search for hyperparametes the runtime is much worse and this overrides previous considerations about training and test time. The MVN and kNN again work well without hyperparameter search, but it is beneficial for MF-LS and absolutely necessary for MF-BPR.

\subsection{Modified MVN, MF-LS, kNN models}


\subsubsection{Missing vs. observed non-interactions}
Based on the mathematical analysis of the methods we hypothesize that MVN has a better predictive accuracy because it regards the non-interactions as missing at prediction time. To test this theory we can apply both the standard and the modified models. Results for the Movie Lens 1M implicit data set are displayed in Table~\ref{ml-1m-mfls}. We see that considering non-interactions as missing significantly improves the predictions for implicit data, and the MF-LS model then reaches the performance of the MVN model. The 'implicit' package we used for kNN in the previous experiments already makes use of this trick because the normalization factor is not included, and accuracy without this trick would be much worse. These results suggest that the same trick should be used with the MF-LS model. 

Increasing the number of latent dimensions $d$ and $k$ is helpful in the models that predict with missing non-interactions as the performance keeps increasing for more latent dimensions. In fact, we can simply set the number of latent factors $d$ in MF-LS and the number of neighbours $k$ in kNN to the number of items because this hyperparameter is not needed to regularize the model. There was no practical benefit to limiting the number of eigenvectors $d$ in MVN, and the number could also be set to the number of items. This is highly beneficial as not having to search for hyperparameters saves a lot of computational time and the model is more robust to not setting the optimal values.


\begin{table}[htbp]
\caption{Implicit Movie Lens 1M: Modified and Standard MVN/MF-LS}
\label{ml-1m-mfls}
\begin{center}
\begin{tabular}{|r|r|r|r|r|}
\hline
\textbf{Non-interactions} & 
\multicolumn{2}{c|}{Missing} & \multicolumn{2}{c|}{Observed} \\
\textbf{Model} & \textbf{Prc@20} & \textbf{nDCG@m} & \textbf{Prc@20} & \textbf{nDCG@m}\\
\hline
MVN (d=m) & 0.504 & 0.722 & 0.449 & 0.681 \\
\hline
MVN (d=256) & 0.505 & 0.723 & 0.448 & 0.680 \\
\hline
kNN (k=m) & 0.481 & 0.720 & 0.011 & 0.574 \\
\hline
kNN (k=256) & 0.406	& 0.682 & 0.241	& 0.619 \\
\hline
MF-LS (d=1024) & 0.517 & 0.733 & 0.478 & 0.716 \\
\hline
MF-LS (d=256) & 0.493 & 0.724 & 0.478 & 0.716 \\
\hline
\end{tabular}
\label{tab1}
\end{center}
\end{table}

\subsubsection{Number of items used as seed}

\begin{figure}
\centerline{\includegraphics[width=0.90\columnwidth]{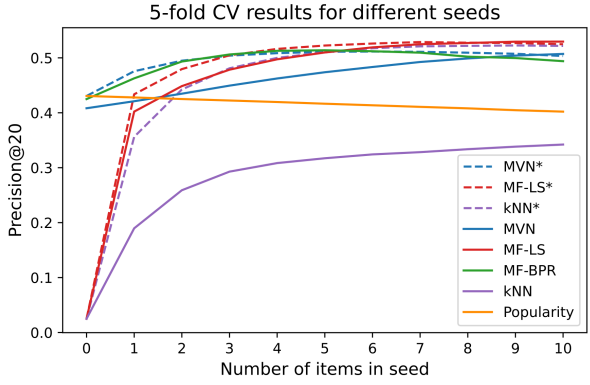}}
\caption{Precision@20 increases with more seed items in the Movie Lens 1M data set and converges at around 5-8 items. The missing non-interactions trick helps the models (MVN*, MF-LS*, kNN*) with small seed sizes, where the ranking based BPR also does well. However, eventually the carefully regularized standard MF-LS and the leave-item-out MVN models reach similar performance as the non-interactions may become informative for large seed sizes, but the standard kNN remains uncompetitive. }
\label{figure:seed}
\end{figure}

Collaborative filtering recommends new items based on the items that the user has interacted with. An important practical aspect is the number of items 'user has interacted with' required as a seed. Fewer items are generally better because then we can give recommendations to more users. The Netflix competition predicted missing movie ratings with a large seed \cite{netflix}, but what is considered as important  has evolved over time. Netflix today (2020) asks for '3 liked movies' when the user first logs in and predicts a list of recommendations. This is the setting we investigated in this paper. Additionally, when a user clicks on a movie, they can see related movies and series - a seed size of only 1. 

To test how many movies are required as a seed in the Movie Lens 1M data set, we did the same evaluation as before but varied the seed size $0,1, ... 10$ in the test set. The results of 5-fold cross evaluation are displayed in Figure~\ref{figure:seed}. The MVN* and BPR models with item biases predict the 'Popularity' as a baseline when the seed size is 0 (no movies given) and from there reach their best precision quite fast with small seed sizes. The modified baselines MF-LS* and kNN* that consider non-interactions as missing become competitive as the seed size grows. The standard MF-LS model and the MVN model with observed non-interactions eventually reach similar performance, but the standard kNN remains substantially worse. 

It is probably possible to improve both the baselines and the MVN based on this observation. The baselines could benefit from including an item biases for small seed sizes, so that their predictions would start at the well performing popularity ranking and user/item interactions would be modelled on top of it. The MVN could benefit from regularizing the item bias to achieve a better trade-off between popularity and user/item interactions. In any case, the models converge at around 5 to 8 seed items and additional items do not help to make better recommendations.

\section{Qualitative results}

\subsection{Example of 'missing' vs. 'observed' non-interactions}

To illustrate the intuition behind why non-interactions should be considered missing at prediction time, consider the following example. For random users and one test user (red) in the Movie Lens 1M data set, we illustrate the implicit data matrix $\interactions$ and co-occurrence matrix $\mathbf{F}=\frac{1}{n} \interactions^T \interactions$ for (Terminator 2: Judgement Day, Toy Story, The Terminator):

\[
\interactions = 
\left( \begin{array}{ccc}
...& ...& ...\\
1& 0& 1\\
1& 1& 1\\
1& 1& 1\\
1& 0& 1\\
0& 1& 0\\
1& 0& 1\\
1& 1& 1\\
0& 0& 0\\
\textcolor{red}{1} & \textcolor{red}{0}& \textcolor{red}{0}
\end{array}\right)
\mathbf{F}=
\left( \begin{array}{ccc}
0.44& 0.21& 0.28\\
0.21& 0.34& 0.17\\
0.28& 0.17& 0.35\\
\end{array}\right)
\]

We see from $\mathbf{F}$ that there is some correlation between the implicit watching status of movies: more user have watched "The Terminator" and the "Terminator 2: Judgement Day" (28\%) than have watched "Toy Story" and "Terminator 2: Judgement Day" (21\%), even though "The Terminator" and "Toy Story" are equally popular movies (34\% vs. 35\%). The test user has watched the movie "Terminator 2: Judgement Day" but has not watched the movie "Toy Story" or "The Terminator". Which one of these should we recommend? 

\begin{figure}
\centerline{\includegraphics[width=\columnwidth]{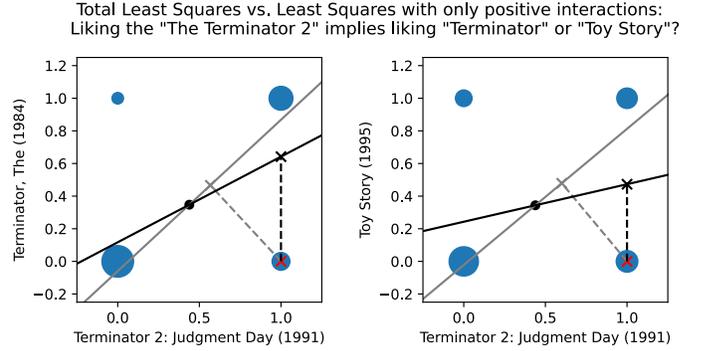}}
\caption{Illustration of how the methods arrive at different predictions for implicit data: SVD (grey) uses the full liked/not-liked vector (1,0) and projects to the first principal component, whereas MVN (black) uses the liked vector (1,\textit{na}) and predicts the expected value of \textit{na}. The predicted regression lines are calculated from the co-occurance matrix (blue dots) and intersect at the mean value (black dot). Based on $(1,0)$, the predictions are $(0.60, 0.48)$ for (Terminator 2, Toy Story) and $(0.57, 0.47)$ for (Terminator 2, The Terminator). Based on $(1,\textit{na})$, the predictions are $(1.00,0.47)$ for (Terminator 2, Toy Story) and $(1.00,0.64)$ for (Terminator 2, The Terminator).}
\label{figure:implicitTLS}
\end{figure}

We illustrate how the correlations can be used to predict the missing watching status. To simplify the visualization we considered two separate prediction tasks: 
\begin{enumerate}
    \item Predict watching 'Toy Story' from 'Terminator 2' (predict column 2 from column 1).
    \item Predict watching 'The Terminator' from 'Terminator 2' (predict column 3 from column 1).
\end{enumerate}
The test user has watched 'Terminator 2' but has not watched the other movie 'Toy Story' or 'The Terminator'. In Figure~\ref{figure:implicitTLS} we visualize how the SVD and the MVN arrive at different predictions. The SVD considers the predicted movie as an explicit '0' whereas the MVN considers it a missing value '\textit{na}'. The SVD ranks 'The Terminator' (0.47) lower than 'Toy Story' (0.48). The MVN ranks the related movie 'The Terminator' (0.64) much higher than 'Toy Story' (0.47). We think that same happens in higher dimensional spaces: the zeros contribute to recommending less related movies.

\subsection{Accuracy vs. Subjective quality}

The ultimate goal of recommender systems is not to maximize accuracy metrics but to produce useful recommendations. The underlying assumption is that more accurate methods produce more useful recommendations. Some doubt is cast on this assumption by the fact that the baseline of recommending the most popular items to every user does so well. Users surely do not see value in a recommender system that always predicts the same popular movies regardless of what movies they like. 
Luckily, we can remove the 'popularity bias' from the MVN method: we set the mean vector to zero and substitute the covariance matrix by the correlation matrix. In the kNN method, we can replace the cosine similarity by the correlation matrix. In the BPR and similar methods, we can omit the item bias terms. The quantitative results are displayed in Table~\ref{without-bias}. 

\begin{table}[htbp]
\caption{Precision@20 in Movie Lens 1M}
\label{without-bias}
\begin{center}
\begin{tabular}{|r|r|r|}
\hline
\textbf{Model} & \textbf{Item bias} & \textbf{No item bias} \\
\hline
Baseline & 0.421 & 0.045 \\
\hline
MVN & 0.503 & 0.412 \\
\hline
kNN & 0.482 & 0.415 \\
\hline
BPR & 0.506 & 0.219 \\
\hline
\end{tabular}
\label{tab1}
\end{center}
\end{table}

We see that the accuracy decreases significantly without the influence of item popularity, it can even be worse than the baseline of recommending the most popular items. However, the qualitative results are a different story. Consider an example user who likes dystopian science fiction and has watched "Alien", "The Terminator", and "The Matrix". Table~\ref{without-bias-recommendations} presents the Top-20 recommendations they receive from the different models with and without item popularity. It looks like the 'more accurate' methods recommend more popular but less related items, and can even recommend very popular but completely unrelated items like "The Princess Bride". However, the model based recommendations without popularity are much better and include other dystopian scifi titles like "Robocop", "Max Max", "The Fifth Element", "Planet of the Apes", etc.

There is no straightforward solution to this problem. One option would be to set up user studies but this can be difficult and expensive, and the results may still be subjective. Measuring more objective things like click-through rates or revenues provides no guarantee that the metric correlates with how useful users felt the recommendations were. The hope with accuracy metrics could be that any improvement above baseline popularity signifies finding a signal based on user and item interactions. If the method then allows us to remove the popularity component, this should mean that more accurate model predicts more user and item interactions.

\begin{table*}
\caption{Given a seed of 'Alien (1979)', 'Terminator, The (1984)', 'Matrix, The (1999)': Recommendations@20}
\label{without-bias-recommendations}
\begin{center}
\begin{tabularx}{\textwidth}{|l|X|X|}
\hline
\textbf{Model} & \textbf{Item bias} & \textbf{No item bias} \\
\hline
Baseline
& 
        'American Beauty (1999)', 'Star Wars: Episode IV - A New Hope (1977)',
       'Star Wars: Episode V - The Empire Strikes Back (1980)',
       'Star Wars: Episode VI - Return of the Jedi (1983)',
       'Jurassic Park (1993)', 'Saving Private Ryan (1998)',
       'Terminator 2: Judgment Day (1991)', 'Matrix, The (1999)',
       'Back to the Future (1985)', 'Silence of the Lambs, The (1991)',
       'Men in Black (1997)', 'Raiders of the Lost Ark (1981)', 'Fargo (1996)',
       'Sixth Sense, The (1999)', 'Braveheart (1995)',
       'Shakespeare in Love (1998)', 'Princess Bride, The (1987)',
       'Schindler's List (1993)', 'L.A. Confidential (1997)',
       'Groundhog Day (1993)'
&
        'Thomas and the Magic Railroad (2000)', 'Back Stage (2000)',
       'Project Moon Base (1953)',
       'Abbott and Costello Meet Frankenstein (1948)',
       'Unbearable Lightness of Being, The (1988)', 'Flamingo Kid, The (1984)',
       'Brown's Requiem (1998)', 'Liberty Heights (1999)',
       'Independence Day (ID4) (1996)', 'Autumn Sonata (Höstsonaten ) (1978)',
       'In Dreams (1999)', 'Two Family House (2000)', 'Matrix, The (1999)',
       'Matewan (1987)', 'Larger Than Life (1996)',
       'Weekend at Bernie's (1989)', 'My Fair Lady (1964)',
       'Fun and Fancy Free (1947)', 'Father of the Bride (1950)',
       'Dark Crystal, The (1982)'
\\
\hline
MVN  & 
'Matrix, The (1999)', 'Terminator, The (1984)', 'Alien (1979)',
       'Star Wars: Episode IV - A New Hope (1977)',
       'Star Wars: Episode V - The Empire Strikes Back (1980)',
       'Terminator 2: Judgment Day (1991)',
       'Star Wars: Episode VI - Return of the Jedi (1983)', 'Aliens (1986)',
       'Raiders of the Lost Ark (1981)', 'Jurassic Park (1993)',
       'Back to the Future (1985)', 'Men in Black (1997)',
       'Total Recall (1990)', 'E.T. the Extra-Terrestrial (1982)',
       'Saving Private Ryan (1998)', 'Blade Runner (1982)', 'Die Hard (1988)',
       'Princess Bride, The (1987)', 'Fugitive, The (1993)',
       'Star Wars: Episode I - The Phantom Menace (1999)'
& 
'Matrix, The (1999)', 'Terminator, The (1984)', 'Alien (1979)',
       'Aliens (1986)', 'Terminator 2: Judgment Day (1991)', 'Predator (1987)',
       'Total Recall (1990)', 'Star Wars: Episode IV - A New Hope (1977)',
       'Star Wars: Episode V - The Empire Strikes Back (1980)',
       'Die Hard (1988)', 'Robocop (1987)', 'Blade Runner (1982)',
       'Star Trek: The Wrath of Khan (1982)', 'Superman (1978)',
       'Mad Max (1979)', 'Indiana Jones and the Last Crusade (1989)',
       'Fifth Element, The (1997)', 'Jaws (1975)', 'Planet of the Apes (1968)',
       'Close Encounters of the Third Kind (1977)'
\\
\hline
kNN & 
'Terminator, The (1984)', 'Alien (1979)', 'Matrix, The (1999)',
       'Aliens (1986)', 'Terminator 2: Judgment Day (1991)',
       'Star Wars: Episode IV - A New Hope (1977)',
       'Star Wars: Episode V - The Empire Strikes Back (1980)',
       'Total Recall (1990)',
       'Star Wars: Episode VI - Return of the Jedi (1983)',
       'Raiders of the Lost Ark (1981)', 'Die Hard (1988)',
       'Men in Black (1997)', 'Jurassic Park (1993)', 'Predator (1987)',
       'Blade Runner (1982)', 'Back to the Future (1985)',
       'Fugitive, The (1993)', 'Indiana Jones and the Last Crusade (1989)',
       'E.T. the Extra-Terrestrial (1982)', 'Robocop (1987)'
& 
'Terminator, The (1984)', 'Alien (1979)', 'Matrix, The (1999)',
       'Aliens (1986)', 'Terminator 2: Judgment Day (1991)',
       'Total Recall (1990)', 'Predator (1987)',
       'Star Wars: Episode IV - A New Hope (1977)',
       'Star Wars: Episode V - The Empire Strikes Back (1980)',
       'Die Hard (1988)', 'Robocop (1987)', 'Blade Runner (1982)',
       'Star Trek: The Wrath of Khan (1982)', 'Fifth Element, The (1997)',
       'Superman (1978)', 'Indiana Jones and the Last Crusade (1989)',
       'Mad Max (1979)', 'Face/Off (1997)', 'Planet of the Apes (1968)',
       'Star Wars: Episode VI - Return of the Jedi (1983)'
\\
\hline
BPR &
'Star Wars: Episode IV - A New Hope (1977)',
       'Star Wars: Episode V - The Empire Strikes Back (1980)',
       'Matrix, The (1999)', 'Terminator 2: Judgment Day (1991)',
       'Alien (1979)', 'Terminator, The (1984)',
       'Star Wars: Episode VI - Return of the Jedi (1983)', 'Aliens (1986)',
       'Total Recall (1990)', 'Raiders of the Lost Ark (1981)',
       'Jurassic Park (1993)', 'Men in Black (1997)',
       'Back to the Future (1985)', '2001: A Space Odyssey (1968)',
       'Blade Runner (1982)',
       'Star Wars: Episode I - The Phantom Menace (1999)',
       'E.T. the Extra-Terrestrial (1982)', 'Abyss, The (1989)',
       'Predator (1987)', 'Jaws (1975)'
&
'Terminator, The (1984)', 'Alien (1979)', 'Aliens (1986)',
       'Predator (1987)', 'Matrix, The (1999)',
       'Terminator 2: Judgment Day (1991)', 'Total Recall (1990)',
       'Mad Max (1979)', 'Robocop (1987)', 'Blade Runner (1982)',
       'Star Wars: Episode IV - A New Hope (1977)', 'Die Hard (1988)',
       'Thing, The (1982)', 'Mad Max 2 (a.k.a. The Road Warrior) (1981)',
       'Fifth Element, The (1997)', 'Star Trek: The Wrath of Khan (1982)',
       'Star Trek IV: The Voyage Home (1986)', 'Starship Troopers (1997)',
       'Star Wars: Episode V - The Empire Strikes Back (1980)',
       'Superman (1978)'
\\
\hline
\end{tabularx}
\label{tab1}
\end{center}
\end{table*}

\section{Conclusion}

We presented a simple new collaborative filtering algorithm for Top-N recommendation in implicit data sets motivated by the Multivariate Normal Distribution (MVN), where the predicted ranking of non-interacted items can be calculated directly from data with a closed form expression. We showed that the method achieves the best ranking accuracy in many public data sets. As an additional benefit, the method is very robust as there is only one regularization hyperparameter which works well when set to zero. The main idea and difference to standard baselines is that the non-interactions are not taken into account when the interactions of every user are predicted. Methods that have been modified as such also have a better accuracy with small seed sizes and exhibit similar robustness to less than optimal hyperparameter choices.


\begin{thebibliography}{00}

\bibitem{review1} Park, D. H., Kim, H. K., Choi, I. Y., \& Kim, J. K. (2012). A literature review and classification of recommender systems research. Expert systems with applications, 39(11), 10059-10072.

\bibitem{review2} Lu, J., Wu, D., Mao, M., Wang, W., \& Zhang, G. (2015). Recommender system application developments: a survey. Decision Support Systems, 74, 12-32.

\bibitem{ecommerence} Schafer, J. B., Konstan, J., \& Riedl, J. (1999, November). Recommender systems in e-commerce. In Proceedings of the 1st ACM conference on Electronic commerce (pp. 158-166).

\bibitem{handbook} Ricci, F., Rokach, L., \& Shapira, B. (2011). Introduction to recommender systems handbook. In Recommender systems handbook (pp. 1-35). Springer, Boston, MA.

\bibitem{topn0} Cremonesi, Paolo, Yehuda Koren, and Roberto Turrin. "Performance of recommender algorithms on top-n recommendation tasks." Proceedings of the fourth ACM conference on Recommender systems. ACM, 2010.

\bibitem{topn1} Deshpande, M., \& Karypis, G. (2004). Item-based top-n recommendation algorithms. ACM Transactions on Information Systems (TOIS), 22(1), 143-177.

\bibitem{topn2} Hu, Yifan, Yehuda Koren, and Chris Volinsky. "Collaborative filtering for implicit feedback datasets." 2008 Eighth IEEE International Conference on Data Mining. Ieee, 2008.

\bibitem{topn3} Nikolakopoulos, A. N., Kalantzis, V., Gallopoulos, E., \& Garofalakis, J. D. (2019). EigenRec: generalizing PureSVD for effective and efficient top-N recommendations. Knowledge and Information Systems, 58(1), 59-81.

\bibitem{mostrelevant} Desrosiers, C., \& Karypis, G. (2011). A comprehensive survey of neighborhood-based recommendation methods. In Recommender systems handbook (pp. 107-144). Springer, Boston, MA.

\bibitem{evaluation1} Herlocker, Jonathan L., et al. "Evaluating collaborative filtering recommender systems." ACM Transactions on Information Systems (TOIS) 22.1 (2004): 5-53.

\bibitem{collaborative} Pazzani, M. J., \& Billsus, D. (2007). Content-based recommendation systems. In The adaptive web (pp. 325-341). Springer, Berlin, Heidelberg.

\bibitem{content} Schafer, J. B., Frankowski, D., Herlocker, J., \& Sen, S. (2007). Collaborative filtering recommender systems. In The adaptive web (pp. 291-324). Springer, Berlin, Heidelberg.

\bibitem{hybrid} Burke, R. (2007). Hybrid web recommender systems. In The adaptive web (pp. 377-408). Springer, Berlin, Heidelberg.

\bibitem{better1} Koren, Y. (2008, August). Factorization meets the neighborhood: a multifaceted collaborative filtering model. In Proceedings of the 14th ACM SIGKDD international conference on Knowledge discovery and data mining (pp. 426-434).

\bibitem{better2} Takács, G., Pilászy, I., Németh, B., \& Tikk, D. (2007). Major components of the gravity recommendation system. Acm Sigkdd Explorations Newsletter, 9(2), 80-83.

\bibitem{coldstart} Bobadilla, J., Ortega, F., Hernando, A., \& Gutiérrez, A. (2013). Recommender systems survey. Knowledge-based systems, 46, 109-132.

\bibitem{evaluation2} Pu, Pearl, Li Chen, and Rong Hu. "Evaluating recommender systems from the user’s perspective: survey of the state of the art." User Modeling and User-Adapted Interaction 22.4-5 (2012): 317-355

\bibitem{evaluation3}  Konstan, Joseph A., and John Riedl. "Recommender systems: from algorithms to user experience." User modeling and user-adapted interaction 22.1-2 (2012): 101-123.

\bibitem{evaluation4} Ziegler, C. N., McNee, S. M., Konstan, J. A., \& Lausen, G. (2005, May). Improving recommendation lists through topic diversification. In Proceedings of the 14th international conference on World Wide Web (pp. 22-32).


\bibitem{explicit_advances} Adomavicius, G., Tuzhilin, A.: Toward the next generation of recommender systems: A survey of the state-of-the-art and possible extensions. IEEE Transactions on Knowledge and Data Engineering 17(6), 734–749 (2005) 


\bibitem{nn_user} Herlocker, J. L., Konstan, J. A., Borchers, A., \& Riedl, J. (2017, August). An algorithmic framework for performing collaborative filtering. In ACM SIGIR Forum (Vol. 51, No. 2, pp. 227-234). New York, NY, USA: ACM.

\bibitem{nn_item1} Sarwar, B., Karypis, G., Konstan, J., \& Riedl, J. (2001, April). Item-based collaborative filtering recommendation algorithms. In Proceedings of the 10th international conference on World Wide Web (pp. 285-295).

\bibitem{nn_item2} Linden, G., Smith, B., \& York, J. (2003). Amazon. com recommendations: Item-to-item collaborative filtering. IEEE Internet computing, 7(1), 76-80.

\bibitem{nn_better} Bell, R. M., \& Koren, Y. (2007, October). Scalable collaborative filtering with jointly derived neighborhood interpolation weights. In Seventh IEEE International Conference on Data Mining (ICDM 2007) (pp. 43-52). IEEE.

\bibitem{latent1} Blei, D. M., Ng, A. Y., \& Jordan, M. I. (2003). Latent dirichlet allocation. Journal of machine Learning research, 3(Jan), 993-1022.

\bibitem{latent2} Hofmann, T. (2004). Latent semantic models for collaborative filtering. ACM Transactions on Information Systems (TOIS), 22(1), 89-115.

\bibitem{latent3} Salakhutdinov, R., Mnih, A., \& Hinton, G. (2007, June). Restricted Boltzmann machines for collaborative filtering. In Proceedings of the 24th international conference on Machine learning (pp. 791-798).

\bibitem{latent4} He, X., Liao, L., Zhang, H., Nie, L., Hu, X., \& Chua, T. S. (2017, April). Neural collaborative filtering. In Proceedings of the 26th international conference on world wide web (pp. 173-182).

\bibitem{whysvd1} Koren, Y., Bell, R., \& Volinsky, C. (2009). Matrix factorization techniques for recommender systems. Computer, 42(8), 30-37.

\bibitem{whysvd2} Koren, Y., \& Bell, R. (2015). Advances in collaborative filtering. In Recommender systems handbook (pp. 77-118). Springer, Boston, MA.

\bibitem{earlysvd} Sarwar, B., Karypis, G., Konstan, J., \& Riedl, J. (2000). Application of dimensionality reduction in recommender system-a case study. Minnesota Univ Minneapolis Dept of Computer Science.

\bibitem{svd1} Canny, J. (2002, August). Collaborative filtering with privacy via factor analysis. In Proceedings of the 25th annual international ACM SIGIR conference on Research and development in information retrieval (pp. 238-245).

\bibitem{svd2} Funk, S. (2006). Netflix update: Try this at home. http://sifter.org/˜simon/journal/20061211.html

\bibitem{svd3} Bell, R., Koren, Y., \& Volinsky, C. (2007, August). Modeling relationships at multiple scales to improve accuracy of large recommender systems. In Proceedings of the 13th ACM SIGKDD international conference on Knowledge discovery and data mining (pp. 95-104).

\bibitem{svd4} Paterek, A. (2007, August). Improving regularized singular value decomposition for collaborative filtering. In Proceedings of KDD cup and workshop (Vol. 2007, pp. 5-8).

\bibitem{svd5} Koren, Y. (2008, August). Factorization meets the neighborhood: a multifaceted collaborative filtering model. In Proceedings of the 14th ACM SIGKDD international conference on Knowledge discovery and data mining (pp. 426-434).

\bibitem{implicit_abundant} Oard, D. W., \& Kim, J. (1998, July). Implicit feedback for recommender systems. In Proceedings of the AAAI workshop on recommender systems (Vol. 83). WoUongong.

\bibitem{implicit_benefit1} Konstan, J. A., Miller, B. N., Maltz, D., Herlocker, J. L., Gordon, L. R., \& Riedl, J. (1997). GroupLens: applying collaborative filtering to Usenet news. Communications of the ACM, 40(3), 77-87.

\bibitem{implicit_benefit2} Terveen, L., Hill, W., Amento, B., McDonald, D., \& Creter, J. (1997). PHOAKS: A system for sharing recommendations. Communications of the ACM, 40(3), 59-62.

\bibitem{implicit_correlation} Marlin, B., Zemel, R. S., Roweis, S., \& Slaney, M. (2012). Collaborative filtering and the missing at random assumption. arXiv preprint arXiv:1206.5267.


\bibitem{learningtorank} Liu, T. Y. (2011). Learning to rank for information retrieval. Springer Science \& Business Media.

\bibitem{preference1} Cohen, W. W., Schapire, R. E., \& Singer, Y. (1999). Learning to order things. Journal of artificial intelligence research, 10, 243-270.

\bibitem{preference2} Jin, R., Si, L., Zhai, C., \& Callan, J. (2003, November). Collaborative filtering with decoupled models for preferences and ratings. In Proceedings of the twelfth international conference on Information and knowledge management (pp. 309-316).

\bibitem{bpr} Rendle, Steffen, et al. "BPR: Bayesian personalized ranking from implicit feedback." Proceedings of the twenty-fifth conference on uncertainty in artificial intelligence. AUAI Press, 2009.

\bibitem{log} Johnson, Christopher C. "Logistic matrix factorization for implicit feedback data." Advances in Neural Information Processing Systems 27 (2014).


\bibitem{lightfm} Kula, Maciej. "Metadata embeddings for user and item cold-start recommendations." arXiv preprint arXiv:1507.08439 (2015).

\bibitem{fm} Rendle, Steffen. "Factorization machines." 2010 IEEE International Conference on Data Mining. IEEE, 2010.



\bibitem{compare1} Rendle, S., Zhang, L., \& Koren, Y. (2019). On the difficulty of evaluating baselines: A study on recommender systems. arXiv preprint arXiv:1905.01395.

\bibitem{compare2} Said, A., \& Bellogín, A. (2014, October). Comparative recommender system evaluation: benchmarking recommendation frameworks. In Proceedings of the 8th ACM Conference on Recommender systems (pp. 129-136).

\bibitem{popularity} Fleder, D., Hosanagar, K.: Blockbuster culture’s next rise or fall: The impact of recommender systems on sales diversity. Management Science 55(5), 697–712 (2009) 

\bibitem{qualitative1} Cremonesi, Paolo, et al. "Looking for “good” recommendations: A comparative evaluation of recommender systems." IFIP Conference on Human-Computer Interaction. Springer, Berlin, Heidelberg, 2011.

\bibitem{qualitative2} Garcin, Florent, et al. "Offline and online evaluation of news recommender systems at swissinfo. ch." Proceedings of the 8th ACM Conference on Recommender systems. ACM, 2014.

\bibitem{qualitative3} Jannach, Dietmar, and Lukas Lerche. "Offline performance vs. subjective quality experience: a case study in video game recommendation." Proceedings of the Symposium on Applied Computing. ACM, 2017.

\bibitem{business1} Jannach, Dietmar, and Kolja Hegelich. "A case study on the effectiveness of recommendations in the mobile internet." Proceedings of the third ACM conference on Recommender systems. ACM, 2009.

\bibitem{business2} Jannach, Dietmar, and Michael Jugovac. "Measuring the business value of recommender systems." ACM Transactions on Management Information Systems (TMIS) 10.4 (2019): 1-23.


\bibitem{movielens} Harper, F. M., \& Konstan, J. A. (2015). The movielens datasets: History and context. Acm transactions on interactive intelligent systems (tiis), 5(4), 1-19.

\bibitem{hetrec} Cantador, I., Brusilovsky, P., \& Kuflik, T. (2011, October). Second workshop on information heterogeneity and fusion in recommender systems (HetRec2011). In Proceedings of the fifth ACM conference on Recommender systems (pp. 387-388).

\bibitem{bookcrossing} Ziegler, C. N., McNee, S. M., Konstan, J. A., \& Lausen, G. (2005, May). Improving recommendation lists through topic diversification. In Proceedings of the 14th international conference on World Wide Web (pp. 22-32).

\bibitem{jester} Goldberg, K., Roeder, T., Gupta, D., \& Perkins, C. (2001). Eigentaste: A constant time collaborative filtering algorithm. information retrieval, 4(2), 133-151.

\bibitem{netflix} Bennett, J., \& Lanning, S. (2007, August). The netflix prize. In Proceedings of KDD cup and workshop (Vol. 2007, p. 35).




\bibitem{shermanmorrison} Sherman, Jack, and Winifred J. Morrison. "Adjustment of an inverse matrix corresponding to a change in one element of a given matrix." The Annals of Mathematical Statistics 21.1 (1950): 124-127.

\bibitem{shrinkage} Ledoit, Olivier, and Michael Wolf. "A well-conditioned estimator for large-dimensional covariance matrices." Journal of multivariate analysis 88.2 (2004): 365-411.

\bibitem{tlsoverview} Markovsky, Ivan, and Sabine Van Huffel. "Overview of total least-squares methods." Signal processing 87.10 (2007): 2283-2302.



\end{thebibliography}
\end{document}